\documentstyle[aps,version2,preprint]{revtex}    
\begin{document}

\draft

\begin{title}
Finite-frequency optical absorption in 1D conductors\\
and Mott-Hubbard insulators
\end{title}

\author{J. M. P. Carmelo$^{1,2}$,  N. M. R. Peres $^{2}$,
and P. D. Sacramento$^{3}$}
\begin{instit}
$^{1}$ CNLS, Los Alamos National Laboratory, Los Alamos, 
New Mexico 87545
\end{instit}
\begin{instit}
$^{2}$ Department of Physics, University of \'Evora,
Apartado 94, P-7002-554 \'Evora, Portugal
\end{instit}
\begin{instit}
$^{3}$ Departamento de F\'{\i}sica and CFIF, IST,
Avenida Rovisco Pais, P-1049-001 Lisboa, Portugal
\end{instit}
\receipt{5 October 1999}

\begin{abstract}
The frequency-dependent conductivity is studied for the 
one-dimensional Hubbard model, using a selection rule, the 
Bethe ansatz, and symmetries associated with conservation
laws. For densities where the system is metallic the 
absorption spectrum has two contributions, a Drude peak at 
$\omega = 0$ separated by a pseudogap from a broad absorption 
band whose lower edge is characterized by a non-classical 
critical exponent. Our findings shed new light on the ``far 
infrared puzzle'' and other optical properties of metallic 
organic chain compounds.
\end{abstract}
\renewcommand{\baselinestretch}{1.656}   


\narrowtext

Organic chain materials \cite{Vescoli} have a broken symmetry 
ground state (GS) and at low critical temperatures $T_{tr}$ 
undergo transitions to unusual $T>T_{tr}$ states. Recent 
$T>T_{tr}$ optical measurements over a huge frequency range 
\cite{Vescoli,Schwartz} established a pseudogap feature together 
with a zero-frequency mode for the ${(TMTSF)}_2X$ salts ($TMTSF$ 
for tetramethyl tetraselena fulvalene.) These experiments have 
raised several important yet unresolved issues. The first 
\cite{Vescoli} was related to the relevance of calculations 
based on the doped 1D Hubbard model. Although these materials 
have finite interchain hopping integrals $t_b$ and $t_c$ with 
$t_a>>t_b>>t_c$ \cite{Vescoli}, hopping becomes 
less effective as the temperature or the frequency is increased 
and the along-chain hopping $t_a$ dominates.

In this Letter we study the optical conductivity 
of the 1D Hubbard model, $\sigma_1 (\omega) 
= 2\pi D\delta (\omega)+\sigma_1^{reg} (\omega)$, where 
$\sigma_1^{reg} (\omega)$ is defined in Eq. (\ref{cond}) 
below. Schulz \cite{Schulz} used the 
Bethe-ansatz (BA) solution to calculate 
$D$ and to derive the total intensity of finite-frequency 
transitions. Our study refers to all parameter space where 
that intensity is significant. Analytical calculations of 
$\sigma_1^{reg} (\omega)$ have so far been restricted to the 
insulator half filling phase in the limit of large onsite 
repulsion $U$ \cite{Gebhard}. In this case the optical gap 
occurs at $E_{opt} = U-4t_a$, followed by an absorption band 
extending up to $U+4t_a$ \cite{Gebhard,Campbell,Mazumdar}. 
Our study confirms the expectation \cite{Vescoli} that 
calculations based on the doped 1D Hubbard model are relevant 
for the theoretical understanding of recent optical experiments 
\cite{Vescoli,Schwartz}. 
In the metallic phase we find that the finite-frequency 
transitions are limited to a well-defined band above a 
pseudogap (which is actually {\it smallest} 
at half filling). The occurrence of such pseudogap 
agrees with predictions from exact diagonalizations 
which have found practically almost no low-frequency 
($\omega \neq 0$) absorption at the metallic phase
\cite{Horsch}. By combining symmetries associated with 
conservation laws \cite{Carmelo00} with the approach of Ref. 
\cite{Carmelo99}, we are able to derive the exponent for 
the frequency dependence above the absorption edge (and 
below the top edge for half filling). We confirm 
\cite{Vescoli} that the existence of 
a pseudogap in the charge excitations with the 
absence of a gap for spin excitations is an indication of 
charge-spin separation in the metallic state and
that that separation is distinct from that of a 1D 
Tomanaga-Luttinger liquid (TLL), 
in which both excitations are gapless \cite{Vescoli}. We
characterize the nature of the unusual metallic state 
which following Ref. \cite{Vescoli} we call {\it 
doped-Mott-Hubbard liquid} (DMHL). Our study reveals that
in the case of the 1D Hubbard model previous naive 
TLL theoretical studies \cite{Gia} of $\sigma_1^{reg} (\omega)$ 
do not apply to values of $\omega$ larger than the 
pseudogap. In contrast to previous
predictions \cite{Schwartz}, we find that 
the phenomenon of suppressed far infrared conductivity
is of 1D character and follows from a parity selection 
rule which also occurs at higher dimensions.
However, only at 1D is this selection rule effective in 
preventing far infrared conductivity transitions. For instance,
in the 2D case doping away from half filling induces significant
mid-infrared absorption \cite{Horsch,Dagotto}. These
results seem to indicate that although for low temperatures just
above $T_{tr}$ the small interchain coupling might lead
to 2D character for some of the properties of the
${(TMTSF)}_2X$ salts, in what the conductivity along the 
chains is concerned the main role of that coupling is 
equivalent to small doping on the single chains, which 
otherwise remain of 1D character, with the electrons 
confined in each chain.  

The 1D Hubbard model in a chemical potential $\mu$
describes $N$ interacting electrons and can be written as 
$\hat{H}={\hat{H}}_{SO(4)}-\mu [N_a-\hat{N}]$, where
${\hat{H}}_{SO(4)}=\hat{T}+U[\hat{D}-\hat{N}/2]$,  
$\hat{T}=-t_a\sum_{j,\sigma}[c_{j\sigma}^{\dag}c_{j+1\sigma} 
+ h. c.]$ is the ``kinetic energy'', $\hat{D} = \sum_{j}
\hat{n}_{j,\uparrow}\hat{n}_{j,\downarrow}$
measures the number of doubly occupied sites,
$\hat{N}=\sum_{j,\sigma}\hat{n}_{j,\sigma}$,
$c_{j\sigma}^{\dagger}$ and $c_{j\sigma}$ are electron 
operators of spin projection $\sigma $ at site $j=1,...,N_a$, 
and $\hat{n}_{j,\sigma}=c_{j\sigma }^{\dagger }c_{j\sigma }$.  
We choose a density $n=N/N_a$ in the interval 
$0\leq n\leq 1$ with even $N$ and zero magnetization
and define $k_F=\pi n/2$. We use units 
such that $-e=\hbar=1$, where $-e$ is the electronic charge. 
The conductivity can be written as

\begin{equation}
\sigma_1^{reg} (\omega) = {\pi\over N_a}
\sum_{\nu\neq 0}{\vert\langle\nu\vert \hat{J}
\vert 0\rangle\vert^2
\over \omega_{\nu,0}}\delta (\omega -\omega_{\nu,0}) \, ,
\label{cond}
\end{equation}
and is also given by $\sigma_1^{reg} (\omega)\propto
\lim_{k\rightarrow 0}{\omega Im\chi_{\rho} (k,\omega)
\over k^2}$, where $\hat{J} = -it_a\sum_{j,\sigma}\left
[c_{j\sigma}^{\dag }c_{j+1\sigma}-c_{j+1\sigma}^{\dag }
c_{j\sigma}\right]$ is the current operator, the  
summation runs over energy eigenstates, 
$\omega_{\nu,0}=E_{\nu}-E_0$ is the excitation energy, 
$\chi_{\rho} (k,\omega)=-\sum_{\nu\neq 0}
{\vert\langle\nu\vert \hat{n}(k)
\vert 0\rangle\vert^2\, 2\omega_{\nu,0}
\over \omega_{\nu,0}^2-(\omega +i\delta)^2}$ 
is the charge-charge response function, and 
$\hat{n}(k)=\sum_{k',\sigma}c_{k'+k,\sigma}^{\dag}
c_{k',\sigma}$. The formula (\ref{cond}) applies if the 
GS $\vert 0\rangle$ is non-degenerate. This is true for 
an even number of sites $N_a$ if periodic (anti-periodic) 
boundary conditions are used for odd (even) values of 
${N\over 2}$, respectively. The GS is then necessarily an 
eigenstate of the parity operator, $\hat{P}_{\pi}$, which 
moves electrons from sites $j$ to $N_a+1-j$, 
$j=1,...,N_a$, and has eigenvalues $\pm 1$. 
Importantly, in the present model $\hat{P}_{\pi}$ 
commutes with the Hamiltonian and anticommutes with the 
current operator. This implies immediately 
that the states $\vert 0\rangle$ and 
$\hat{J}\vert 0\rangle$ have opposite parities, a {\it key 
selection rule} for optical transitions. The final 
states $\vert\nu\rangle$ can be characterized in 
terms of holons, antiholons, spinons \cite{Essler,Carmelo96}, and 
a charge-transfer band (CTB) \cite{Carmelo97}. The holon/antiholon 
and spinon bands describe low-energy charge and spin 
excitations, whereas the CTB is associated 
with the upper Hubbard band \cite{Schulz}. We use the 
labels $\alpha =c,s,t$ for the holon/antiholon band, the spinon
band, and the CTB, respectively 
\cite{Carmelo97}, and the quantum number $\beta$ for 
distinguishing between holons ($\beta =-{1\over 2}$) and 
antiholons ($\beta =+{1\over 2}$) \cite{Carmelo96,Carmelo97}. 
In the present context, the excitations that couple
to the GS are such that the $s$ band is empty of 
spinons, the $c$ band can be populated by holons and by zero 
or one antiholon, and the $t$ band can have occupancy zero or 
one. The momentum variable has the 
form $q_j={2\pi\over N_a}I^{\alpha}_j$, where 
$I^{\alpha}_j$ are successive integers or half-odd integers. 
In contrast to the case of electronic bands, 
the number of available momenta 
$N^*_{\alpha}$ can change with band occupancies. These 
numbers are $N^*_c=N_a$, $N^*_s=N/2-N_t$, and
$N^*_t=N_a-N+N_t$, where $N_t$ is the number of occupied
momenta in the CTB, and the momentum
band widths are $\Delta q_c=2\pi$, $\Delta q_s=2k_F-
2\pi N_t/N_a$, and $\Delta q_t=2\pi -4k_F+2\pi N_t/N_a$.
The numbers $N^h_{c,-{1\over 2}}$ (holons),  
$N^h_{c,+{1\over 2}}$ (antiholons), and $N_t$ are good
quantum numbers which obey the sum rules \cite{Carmelo97}, 
$N_a-N=-2\sum_{\beta=\pm {1\over 2}}\beta N^h_{c,\beta}
=N^h_c-2N_t$, where $N^h_c=\sum_{\beta=\pm {1\over 2}} 
N^h_{c,\beta}$. The GS is characterized by
$N^h_{c,+{1\over 2}}=N_t=0$ and $N^h_{c,-{1\over 2}}=N_a-N$,
with a symmetrical holon occupancy in the $c$ band
for momenta $2k_F<\vert q\vert <\pi$. The GS
$\alpha$ pseudo-Fermi momenta read 
$q_{F\alpha}=\pi N_{\alpha}/N_a$
where $N_{\alpha}=N^*_{\alpha}-N^h_{\alpha}$
and thus $q_{Fc}=2k_F$, $q_{Fs}=k_F$, and $q_{Ft}=0$. 
The energy bands $\epsilon^0_{\alpha}(q)$
are defined by Eqs. (110)-(112) and (C7)-(C9) of Ref. 
\cite{Carmelo97}. [See also Figs. 7 and 8 of Ref. \cite{Carmelo91} 
for $\epsilon_{\alpha}(q)=\epsilon^0_{\alpha}(q)-
\epsilon^0_{\alpha}(q_{F\alpha})$ and $\alpha=c,s$.]
The velocities $v_{\alpha}(q)=d\epsilon^0_{\alpha}(q)/dq$
and $v_{\alpha }\equiv v_{\alpha}(q_{F\alpha})$
and the mass $1/m^*_{\alpha}\equiv
\vert dv_{\alpha}(q)/dq\vert_{q=q_{F\alpha}}$ 
are also important quantities. In the parameter region 
of appreciable oscillator strength for optical transitions,
which corresponds to $n=1$ and to densities close to $n=1$ 
for large enough (but not too large) values of $U$ 
\cite{Schulz}, the CTB width 
$W_t=\epsilon^0_t(0)-\epsilon^0_t(\pi-2k_F)=\epsilon^0_t(0)$ 
either vanishes ($n=1$) or is very small. There
is nearly no oscillator strength for $U<U^*$, where 
$U^*=U^*(n)$ is such that $E_c(n,U^*)=0$. Here 
$E_c(n,U)=E_c\equiv -2\epsilon^0_c(\pi)$ and 
$E_c>0$ (and $E_c<0$) for $U>U^*$ ($U<U^*$) and 
$E_c\rightarrow U-4t$ ($E_c\rightarrow E_{MH}$) as 
$n\rightarrow 0$ ($n\rightarrow 1$). ($E_{MH}$ is 
the $n=1$ Mott-Hubbard gap \cite{Carmelo97}.) $U^*$ 
is a decreasing function of $n$ given by $U^*=0$ 
($U^*=4t$) for $n\rightarrow 1$ ($n\rightarrow 0$). 

The main transitions contributing to $\sigma_1^{reg} 
(\omega)$ are of two types: (a) those leaving the band filling 
$N^h_{c,-{1\over 2}}$, $N^h_{c,+{1\over 2}}$, and $N_t$ 
unchanged and (b) those changing these numbers by
$\Delta N^h_{c,-{1\over 2}}=\Delta N^h_{c,+{1\over 2}}= 
\Delta N_t=1$. The former start, in principle, at
$\omega=0$, while the latter have an onset at
$E_{opt}=W_t-2\epsilon^0_c(2k_F)>0$. We find below
that for $n\neq 1$ (or $n=1$) $E_{opt}$ is a pseudogap 
(or a gap). Importantly, for $\omega< E_{opt}$ the 
final-state Hilbert subspace is spanned by states with no 
occupancy in the $t$ band. In this case the excitations
are described by holons and spinons and at
low values of $\omega $ and $n\neq 1$ the quantum
liquid is a TLL. On the other hand, for
$\omega >E_{opt}$ the final-state subspace is
spanned by states also involving $t$ occupation,
the charge and spin excitations remaining
separated and described by holons and spinons.
For $n\neq 1$ this $c$, $s$, and $t$ quantum liquid 
describes the DMHL. The collisions involving holons, 
antiholons, spinons, and $t$ particles do not lead to energy
and momentum transfer and only give rise to shifts in 
the phases of these elementary particles. We find below that
for both the TLL and DMHL the critical exponent expressions are 
simply a linear superposition of the phase-shift parameters
$\xi^j_{\alpha\alpha'}=\delta_{\alpha,\alpha'}+ 
\sum_{l=\pm 1}l^{j}\Phi_{\alpha\alpha'}(q_{F\alpha},
lq_{F\alpha'})$ with $j=0,1$. Here 
$\Phi_{\alpha\alpha'}(q,q')$ are the two-particle 
forward-scattering phase shifts defined by Eqs. (103) 
and (B30)-(B38) of Ref. \cite{Carmelo97}. For
holons and spinons the 8 parameters $\xi^j_{\alpha\alpha'}$
read $\xi^1_{cc}=2\xi^1_{cs}=1/\xi^0_{cc}=
\sqrt{2K_{\rho}}$, $\xi^1_{ss}=-\xi^0_{sc}=1/\xi^0_{ss}=
1/\sqrt{2}$, and $\xi^1_{sc}=\xi^0_{cs}=0$ where 
$K_{\rho}$ is the TLL parameter defined in Ref.
\cite{Schulz} and $\xi^1_{\alpha\alpha'}$
are the entries of the transverse of the conformal-field
theory (CFT) dressed-charge matrix \cite{Frahm}. 
The TLL critical expressions involve 
$v_c$, $v_s$, and $K_{\rho}$ \cite{Schulz}.
On the other hand, the low-energy $(\omega-E_{opt})>0$
DMHL critical theory (above the pseudogap)
is a three-component quantum liquid
of the general type studied in Ref. \cite{Carmelo99}
and involves $v_c$, $v_s$, $m^*_t$, and the 18 parameters
$\xi^j_{\alpha\alpha'}$. Importantly, some of these  
parameters are $n$ and $U$ dependent and cannot be 
expressed in terms of the TLL quantity
$K_{\rho}$. (The $n=1$ and $(\omega-E_{MH})>0$ critical theory 
is also of that type and involves $v_{s}$, $m^*_c$, and 
$m^*_t$.)

The transitions of type (a) do not exist at $n=1$ and we show 
first that, as a consequence of the selection rule presented 
above, they have a very small weight for $n<1$. 
These are excitations within the holon band which 
can be characterized in terms of single and 
multiple electron-hole excitations \cite{Carmelo96b}.
Single electron-hole excitations give no 
contributions to the finite-frequency absorption, while 
multiple excitations are expected to decrease very 
rapidly in intensity. For low $\omega$ we evaluate
$\sigma_1^{reg}(\omega)$ from $\chi_{\rho} (k,\omega)
\propto\int dx\int dt e^{i[kx -\omega t]}
\chi_{\rho} (x,t)$, with the charge-charge
correlation function $\chi_{\rho} (x,t)$ obtained
directly from CFT \cite{Frahm}. Double excitations, which 
would give a contribution $\sigma_1(\omega )=C_3\,\omega^3$, 
are forbidden by the parity selection rule and thus 
$C_3=0$, since these transitions require 
symmetrical changes of holon occupancies at $q$ and $-q$, 
leaving the parity unchanged. This explains why the
$\omega^3$ absorption was not observed in optical
experiments \cite{Schwartz}. Higher-order low-energy
processes would give rise to $\sigma_1(\omega )=C_7\,\omega^7$ for
$n\neq 1$ and to $\sigma_1(\omega )=A_j \,\omega^{4j^2K_{\rho}
-1}$ for $n=1/j$ with $j=2,3,4...$ (due to {\it Umklapp} 
processes). [These two exact low-$\omega$ 
CFT expressions can be obtained by multiplying 
the perturbative expressions $\omega^3$ and 
$\omega^{4j^2K_{\rho}-5}$, respectively, of Ref. \cite{Gia} 
by the factor $\omega^4$, which is missing in the latter
two expressions.] Since these processes require multiple 
excitations in the holon band, they have extremely low 
spectral weight and $C_7\, ,A_j\approx 0$. These considerations 
explain why numerical studies yield so weak features at 
low frequencies \cite{Horsch}. While in 1D the 
$\hat{P}_{\pi}$ eigenvalues $\pm 1$ are directly related 
to the right and left state occupancies
and the selection rule is very restrictive, at higher 
dimensions it is less effective because 
there are many angular-momentum channels which lead 
to significant spectral weight at small values of $\omega$ 
\cite{Dagotto}. The weakness of type (a) 
transitions implies that the onset of 
type (b) transitions at $E_{opt}$ represents a pseudogap 
which for $n<1$ and $U>U^*$ looks like a true gap in an actual 
experiment \cite{Vescoli,Schwartz}. $E_{opt}$, shown in Fig. 1
(a), increases with increasing $U$ and decreases with increasing 
density. For $n=1$ it coincides with the Mott-Hubbard gap, 
$E_{MH}=E_c(1,U)$. Since larger interchain transfer integral
$t_b$ means larger effective doping for the single
chains, this theoretical result is fully consistent with 
the experimental data of Fig. 3 of Ref. \cite{Vescoli}.
The important processes for the
type (b) transitions are those where in addition to a 
particle at $q'$ in the $t$ band a holon-antiholon pair 
is created in the $c$ band at $q$ and $q'-q$ 
(with $q'=0$ for $n=1$). This leads to final states whose 
excitation energy has its minimum and 
maximum values at $\omega=E_{opt}$
and $\omega =E_{opt}+\Delta\omega_{opt}$,
respectively. Here $\Delta\omega_{opt} = 8t_a - 2\epsilon^h_F$ 
where $\epsilon^h_F=\epsilon^0_c(\pi)-\epsilon^0_c(2k_F)
=0$ at $n=1$. This spectral range agrees with large $U$ 
expansions \cite{Gebhard} and numerical calculations
\cite{Campbell,Horsch}. For $n=1$ there are only
these final states and for $n<1$ the missing multipair 
processes contain nearly no spectral weight. 
The current operator cannot produce more than one 
doubly occupied site. Since from the 
results of Refs. \cite{Carmelo97,Carmelo96b} we find 
that creating one $t$ particle involves processes 
which create one of these sites, there will be almost 
no spectral weight above the upper edge at $\omega = 
E_{opt}+\Delta\omega_{opt}$. 

The evaluation of the line shape is a very 
difficult problem. We could solve it both for $n\leq 1$ 
and the region immediately above threshold and for
$n=1$ and energies immediately below the top absorption edge.
Since the number operators ${\hat{N}}_{\alpha}$ are 
conservation laws \cite{Carmelo00} and the $t$ band is 
quadratic for small $q$, the low-energy 
spectrum of the {\it Conservation-Laws Hamiltonian}, 
${\hat{H}}_{CL}\equiv {\hat{H}}_{SO(4)}-\sum_{\alpha}
\epsilon^0_{\alpha}(q_{F\alpha}){\hat{N}}_{\alpha}
-\mu N_a$, is of the type studied in Ref. \cite{Carmelo99}.
The use of the non-linear band approach of that reference 
allows the evaluation of asymptotic expressions for the
charge-charge correlation function of ${\hat{H}}_{CL}$, 
$\chi_{\rho}^{CL} (x,t)$. However, that approach {\it does 
not} provide finite-energy correlation-function 
expressions. Fortunately, the following symmetries and 
properties of ${\hat{H}}$ and ${\hat{H}}_{CL}$ allow 
solution of this problem \cite{Carmelo00}: 
(i) the leading term in $\chi_{\rho}^{CL} (x,t)$ has 
contributions only from transitions to a Hilbert subspace with 
fixed $N_{\alpha}$ numbers; (ii) $[{\hat{H}},{\hat{H}}_{CL}]=0$,
and thus ${\hat{H}}$ and ${\hat{H}}_{CL}$ have the same 
eigenstates; and (iii) for the final subspace relevant to the
conductivity, the difference in energies 
is $E_{opt}=\langle\psi\vert{\hat{H}}\vert\psi\rangle -
\langle\psi\vert{\hat{H}}_{CL}\vert\psi\rangle$
and is {\it the same for all final states
$\vert\psi\rangle$} in that subspace. Using these results, 
we can show that the corresponding leading term in the  
Hubbard model correlation function is of the 
form $\chi_{\rho} (x,t)=
e^{iE_{opt} t}\chi_{\rho}^{CL} (x,t)$. This provides
$\chi_{\rho} (k,\omega)=\int dx\int dt e^{i[kx -(\omega -E_{opt})t]}
\chi_{\rho}^{CL} (x,t)$ for small values of both $k$ and 
$(\omega-E_{opt})>0$ which for $n<1$ leads to 
$\sigma_1^{reg}(\omega)= C\Bigl({[\omega -E_{opt}]
\over E_{opt}}\Bigl)^{\zeta }$ at the onset 
and a critical exponent $\zeta = - {3\over 2} + 
{1\over 2}\Bigl(\sqrt{2\over K_{\rho}}-\xi^0_{ct}\Bigl)^2 + 
{1\over 2}\Bigl(\xi^0_{st}\Bigl)^2$. As expected,
this DMHL exponent cannot be expressed in terms of
the TLL parameter $K_{\rho}$ only. It is shown in Fig. 1 (b) 
and approaches ${1\over 2}$ both as 
$n\rightarrow 1$ at finite $U$ and as $U\rightarrow\infty$, 
and it increases with decreasing values of $U$. $C$ 
increases with increasing $n$ and for $U<U^*$ is vanishing 
small. Consistently with the $n<1$ results, at $n=1$ 
we find $\sigma_1^{reg}(\omega)=C\Bigl({[\omega 
-E_{MH}]\over E_{MH}}\Bigl)^{1/2}$ for all finite values of $U$,
in contrast to the naive $\Bigl(\omega 
-E_{MH}\Bigl)^{-1/2}$ prediction of Ref. \cite{Gia}. 
At $n=1$ we could also find that $\sigma_1^{reg}(\omega)$ 
has a maximum at $\omega=\omega_0$ with $E_{MH}<\omega_0
\leq E_{MH}+4t_a$ and $\omega_0\rightarrow E_{MH}+4t_a$
as $U\rightarrow\infty$, and that for energies 
immediately below the top conductivity edge at
$E_{MH}+8t_a$, $\sigma_1^{reg}(\omega)$ reads
$\sigma_1^{reg}(\omega)=C'\Bigl({[E_{MH}+8t_a-\omega]\over 
E_{MH}}\Bigl)^{1/2}$. At $U$ large the ratio $C'/C$ equals 
one and the conductivity absorption is symmetric around 
$\omega=\omega_0=E_{MH}+4t_a\approx U$, in agreement 
with the results of Refs. \cite{Gebhard,Mazumdar}. However,
$C'/C$ decreases with decreasing $U$ and 
$\omega_0$ is shifted to lower values of $\omega$.
In this case $\sigma_1(\omega)$ is not symmetric around 
$\omega_0$ and a dependence $\sigma_1(\omega)\propto 
\omega^{-\gamma}$ is expected for a conductivity descending 
region of intermediate $\omega$ values of the domain 
$\omega\in (\omega_0\, ,E_{opt}+8t_a)$. (In contrast to
the onset exponent $\zeta$, $\gamma$ cannot be calculated 
within our critical theory and the use of TLL 
\cite{Schwartz,Gia} to estimate it is an open question.) 

The $X=ClO_4$ and $T=10K$ curves of Figs. 
1 (B) of Ref. \cite{Vescoli} and Figs. 3 and 4 of Ref. 
\cite{Schwartz} correspond to a $\omega=0$ Drude peak with 
a very small carrier density separated by a pseudogap 
$E_{opt}\approx 0.014\, eV$ from an absorption band 
with $\gamma\approx 1.3$ and a width $\Delta\omega_{opt}
\approx 1.00\, eV$. These $E_{opt}$ and $\Delta\omega_{opt}$
values can be understood in the framework 
of the 1D Hubbard model with effective values $t_a\approx\, 
0.125\, eV$ [tight-binding model (TBM) calculations
\cite{Vescoli,Schwartz} lead to $t_a\approx\, 
0.250\, eV$ because the energy width of the $\epsilon_c(q)$ 
band GS occupancy, $4t_a-\epsilon^h_F$, is about twice 
that of the TBM band] and $U/t_a\approx 1.5$ if the 
system has a nearly half-filled band, 
$n\approx 0.995$, in which case the optical 
threshold $E_{opt}\approx E_{MH}\approx 0.11\, t_a\approx 0.014\, 
eV$ and $\Delta\omega_{opt}\approx 8t_a\approx 1.00\, eV$. 
Moreover, combining the theoretical values for 
$\Delta\omega_{opt}\approx 8t_a$ and the sumrule 
$\int_{-\infty}^{\infty}\sigma_1 (\omega)$ \cite{Schulz},
we find $C\approx 1.2\, \sigma_1^{reg}(\omega_0)\approx
3600\,\Omega^{-1}cm^{-1}$. A crucial text for our theoretical
description is whether the exponent 
$\zeta $ obtained from the observed sharp onset of absorption 
\cite{Vescoli,Schwartz} agrees with the $t_a$, $U$,
and $n$ values extracted from $E_{opt}$ and $\Delta\omega_{opt}$.
Except for the small onset temperature tail, we
find from the above figures the linear
behavior, $\ln \Bigl({\sigma_1^{reg}(\omega)
\over\sigma_1^{reg}(\omega_0)}\Bigl)
\approx\zeta\,\ln \Bigl({[\omega -E_{opt}]\over E_{opt}}\Bigl) + 
\ln ({C\over \sigma_1^{reg}(\omega_0)})$ 
for ${[\omega -E_{opt}]\over E_{opt}}<0.5$, where
${\zeta }\approx 0.65$ and $C\approx 3530\,\Omega^{-1}
cm^{-1}$. The corresponding asymptotic theoretical conductivity
curve is plotted in Fig. 2. Importantly, Fig. 1 (b) reveals 
that this exponent is consistent with the above $U/t_a$ and 
$n$ values.

We thank D. Baeriswyl, D.K. Campbell, E. Dagotto, T. Giamarchi, 
P. Horsch, and J.M.B. Lopes dos Santos for illuminating 
discussions. This research was supported by the Department 
of Energy under contract W-7405-ENG-36 and by PRAXIS 
under Grant No. 2/2.1/FIS/302/94. 



\figure{(a) The pseudogap $E_{opt}$ in units of $t_a$ and 
(b) the exponent $\zeta$ as function of $n$ for different 
values of $U$. The solid (dashed) lines correspond to
$U>U^*$ ($U<U^*$).
\label{fig1}}
\figure{The conductivity $\sigma_1^{reg}(\omega)$ of 
$(TMTSF)_2ClO_4$ for $T=10K$ as a function of $[\omega-
E_{opt}]/E_{opt}>0$ [from Fig. 4 of Ref. \cite{Schwartz}] 
and the theoretical asymptotic curve $C\Bigl([\omega-
E_{opt}]/E_{opt}\Bigl)^{\zeta}$ (solid line) with 
$\zeta =0.65$ and $C=3530\,\Omega^{-1}cm^{-1}$. 
\label{fig2}}
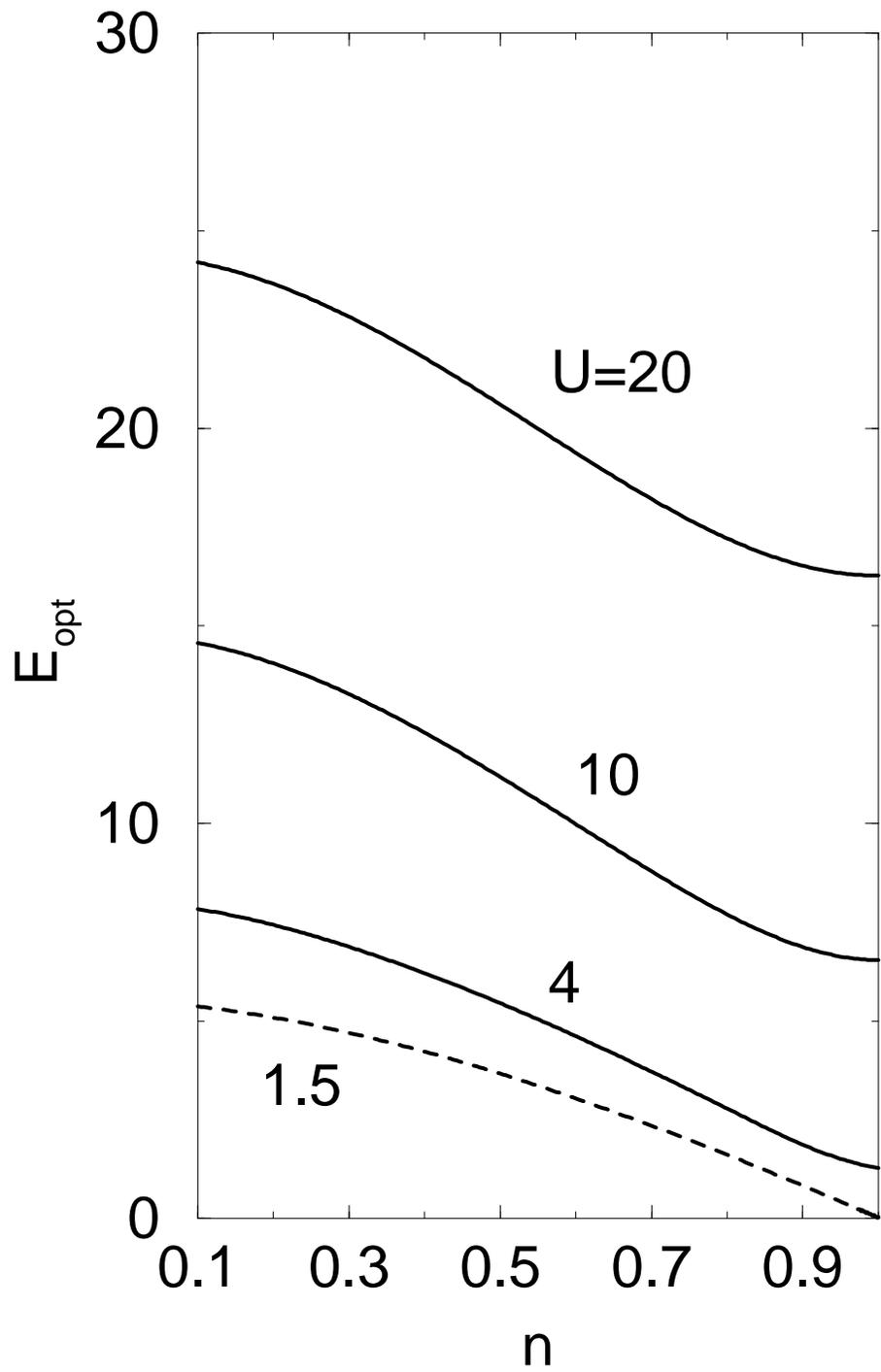
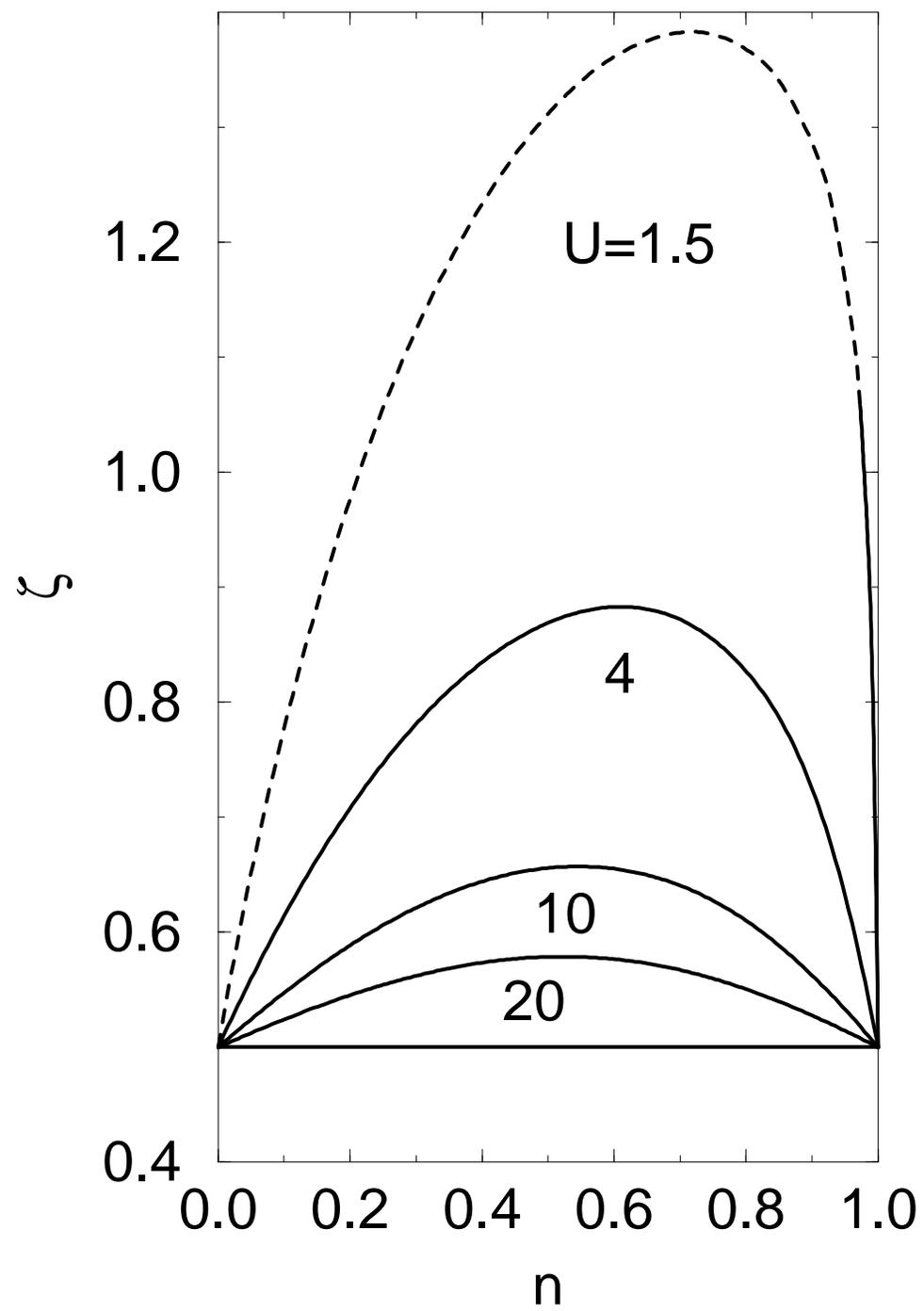
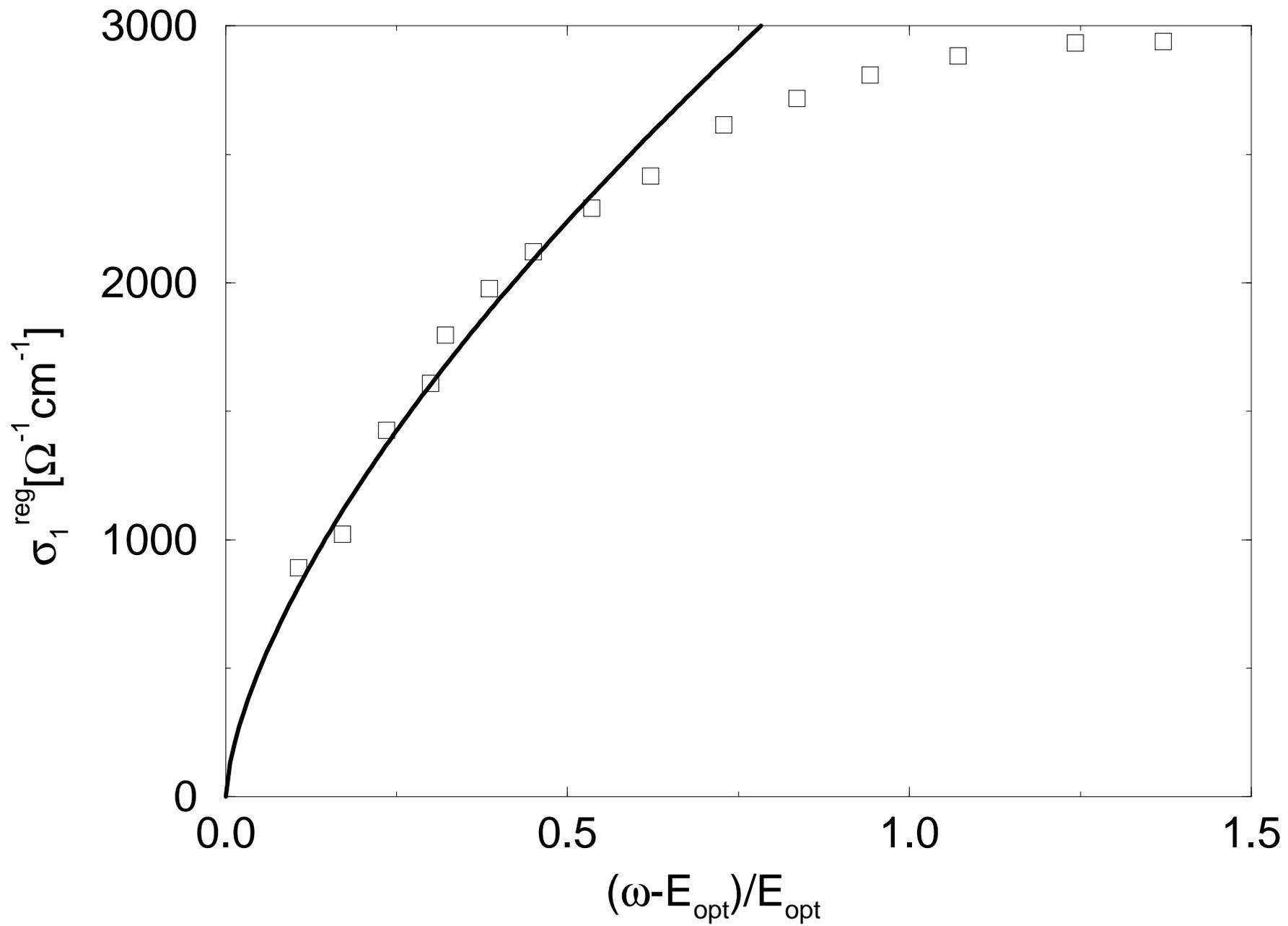
\end{document}